\documentclass[aps,letterpaper,twocolumn]{revtex4}
\usepackage{graphicx}
\usepackage[colorlinks=true, pdfstartview=FitV, linkcolor=blue, citecolor=blue, urlcolor=blue]{hyperref}
\usepackage{dcolumn}% Align table columns on decimal point

\newcommand{\be}{\begin{equation}}
\newcommand{\ee}{\end{equation}}

\newcommand{\bea}{\begin{eqnarray}}
\newcommand{\eea}{\end{eqnarray}}

\begin{document}

\title{
Using Garlic As A Far-Transfer Problem of Proportional And Probabilistic Reasoning
}

\author{Nathan Moore}
\email{nmoore@winona.edu}
\affiliation{
	Department of Physics, 
	Winona State University
	Winona, MN 55987, USA
}
\author{John Deming}
\email{jdeming@winona.edu}
\affiliation{
	Department of Chemistry, 
	Winona State University
	Winona, MN 55987, USA
}
\date{\today}

 \begin{abstract} %
The work describes a general problem, which emphasizes proportional reasoning and probabilistic reasoning skills in the context of planting garlic in a backyard garden.  Along with practicing these reasoning skills in a context far-removed from the standard high school or college curriculum, our solution involves the development of a few relatively sophisticated statistical concepts, specifically histograms and confidence intervals.
\end{abstract} %

\maketitle

\section{Introduction}
In my house, dinner is poorly planned (on a regular basis) and it is embarrassingly common to be halfway through a recipe, with onions sautŽing on the stove, and find that the bulb of garlic I thought was fine a week ago is now sprouting pale green shoots, about $5$ cm long.  In the past I would throw this sprouted garlic away, as once sprouted, the cloves are soft and bitter.  A few springs ago this happened around the time I was planting the vegetable garden, and instead of planting the sprouted bulb in the trash or the compost heap, I planted it next to the lettuce, where it flourished. 

\begin{center}
\begin{figure}
\includegraphics[width=\columnwidth]{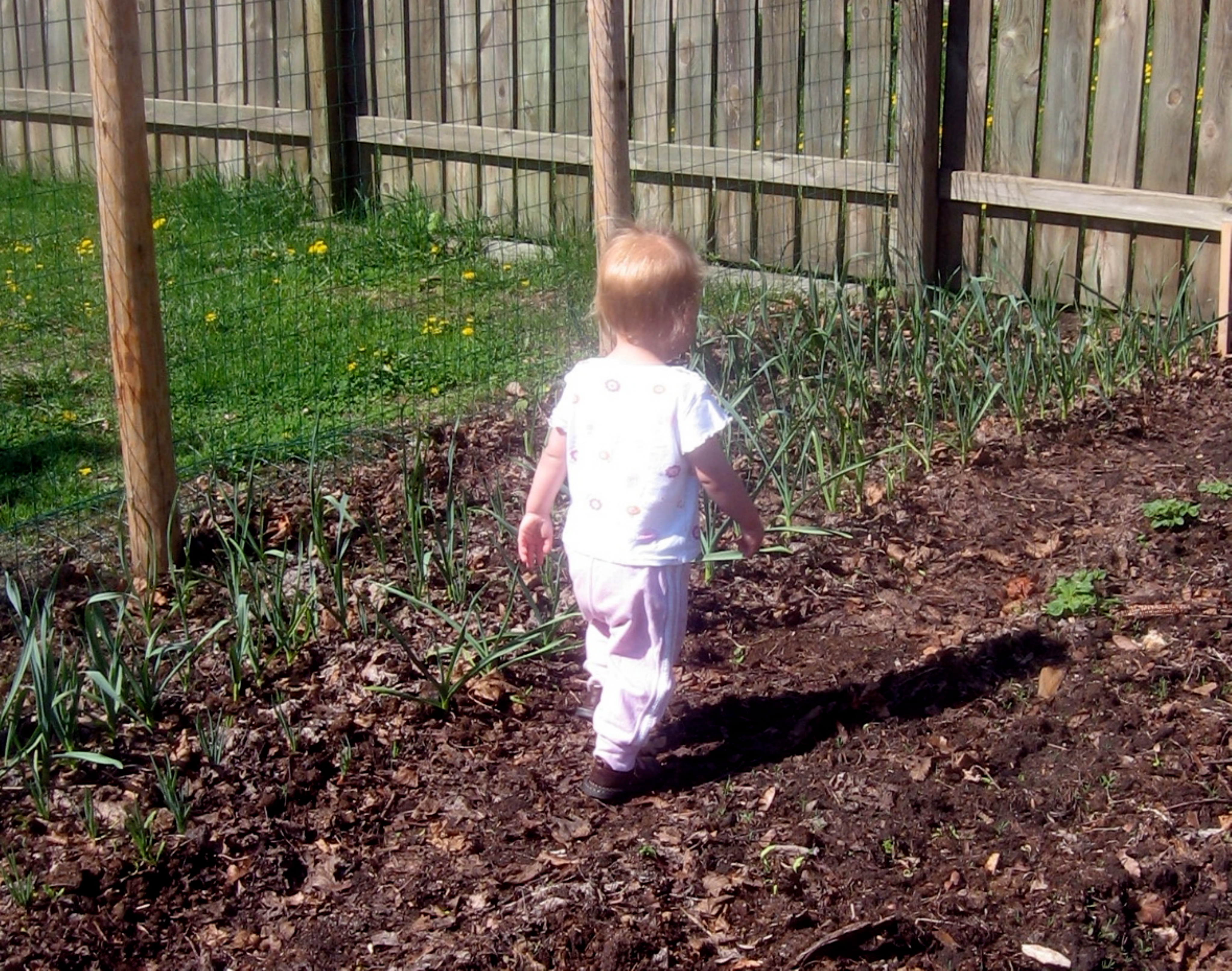}
\caption{Because it is planted in the late fall, Garlic is one of the earliest spring vegetables available.  In this picture, taken in late April $2007$, the shoots produced by individual cloves are well ahead of all other garden vegetables.  For contrast, spring planted potatoes, another vegetable that produces a large amount of foliage through the use of stored energy (starch), are in the right side of the figure.
}
\label{fig:garlic_shoots}
\end{figure}
\end{center}

The problem described in this paper starts from the story above and develops several themes related to dimensional analysis, and also introduces students to a few basic statistical ideas.  The problem was used in a preparatory chemistry class at Winona State University, which in terms of content, is a course designed for students who did not take chemistry in high school. However, this course is unique because one of the primary course goals is the development of students' scientific thinking skills, including those described by Piaget. These skills include control and exclusion of variables, classification, ratio and proportion, probability, combinatorial logic, the ability to formulate and manipulate mental models, and the ability to formulate and systematically test alternative hypotheses against data. Philosophically, the course is organized in a Constructivist framework, and we purposefully choose learning environments that challenge students to develop concepts from data, so the teaching strategies we employ follow the Learning Cycle format\cite{Lawson}. Most of the units were developed for high school chemistry in the ``Frameworks for Inquiry'' program at The University of Montana, as part of a Mathematics and Science Partnership Grant administered by the Montana Office of Public Instruction. Those units follow the historical development of scientific concepts like mass, atoms, thermodynamic properties, etc. The problem described herein enhances those units because it challenges students to apply their knowledge of measurement and proportions in a novel (to the student) context, and provides a place to practice probabilistic reasoning, a hard-to-target skill in physical science classrooms.

In a normal year, garlic is harvested in early to mid summer when the green tops begin to yellow and wilt.  The bulbs are dried for a few weeks and then ideally are stored at $65\%$ humidity, close to $0^{\circ}C$\cite{Coleman}, for use over the following year.  A fraction of this harvest is planted in a well-manured section of the garden in the early fall, where each clove will set roots and perhaps produce a small green shoot.  If properly mulched, garlic will survive the winter without problem and send up green shoots well before anything else in the following spring (see Figure \ref{fig:garlic_shoots} for an illustration).  In Minnesota at least, eating fresh garlic greens while the neighbors are putting away their snowblower for the season is an existential pleasure\cite{Don}.

\begin{center}
\begin{figure}
\includegraphics[width=\columnwidth]{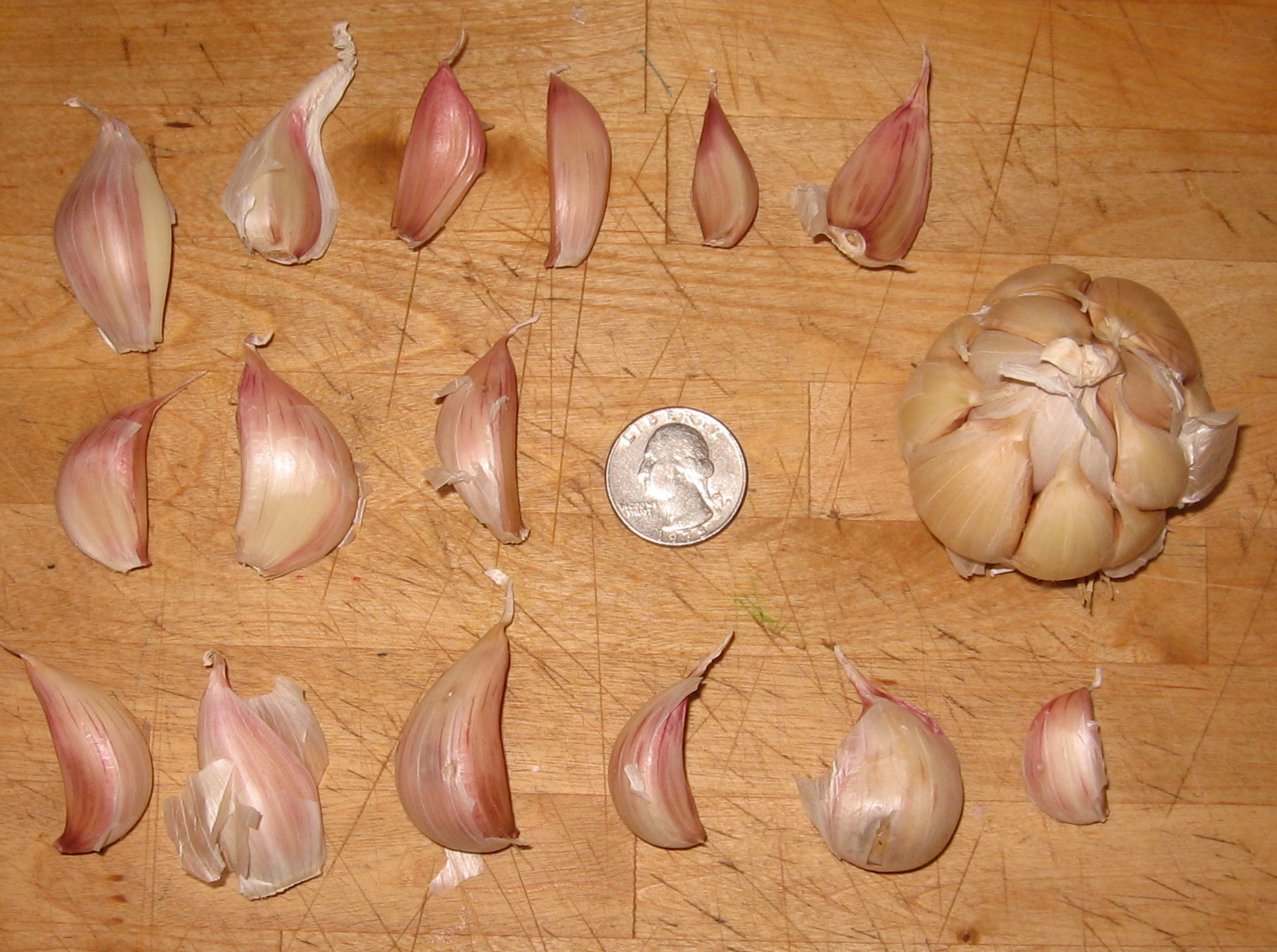}
\caption{
An intact bulb, and a second bulb broken into cloves.  For the bulb broken into cloves, all cloves are probably plantable, although $2$ are near the ``dime'' threshold we used in class.  Additionally, a few of the cloves in the bottom row show bruising, which may inhibit germination in the spring.  An interesting learning cycle that this might initiate would involve the students divide cloves up into ``healthy'', ``diseased", and ``questionable" groups and then plant these groups in a school garden in the fall.  When the cloves sprout in the following spring (near the end of the academic year), students could test their predictions.  Additional predictions that could be tested in a class might include be the germination ratio as a function of clove mass, garlic variety, mulch depth, etc.
}
\label{fig:garlic_cloves}
\end{figure}
\end{center}

Garlic is most easily managed if planted in a rectangular ``patch'' rather than a traditional row.  I plant bulbs 4 inches apart in several dense rows, spaced 6 inches apart.  For more on this sort of dense vegetable cultivation, see \cite{Bartholomew,Adams}.  Over the spring, the plants will first produce green shoots, and later miniature seed heads, called scapes, both of which are tasty and climatically ``early'' vegetables.

The only problem with saving one's own seed after the harvest is that crop failure is a fact of life and, this year, I lost nearly all my seed garlic to mold when a week of rain frustrated my drying rack.  Fortunately for me, the garlic sold in bulk at my local grocery store is quite viable (unlike potatoes, which are chemically treated to reduce the probability of sprouting in the cupboard\cite{Cornell}), and I was able to buy sufficient seed to plant a large patch of cloves in my backyard.

Reluctant to buy too much seed, and seeing this as a concrete example for our students, we gave each student in the class a bulb of garlic ($\approx\$8.00$ for twenty bulbs) and presented the following problem.

Suppose I want to grow enough garlic for the coming year, and have enough left over for seed to plant in the fall:
\begin{enumerate}
\item How much garlic should I purchase to have adequate seed?
\item How big an area do I need to cultivate for this seed?
\item How much money will I save by buying and growing plants, rather than buying bulbs of equivalent quality at the supermarket?
\end{enumerate}
The imaginative instructor will probably realize that these questions can produce lively discussion on a number of relevant topics, most of which can be described quantitatively. In some cases, students may voice questions that can lead to additional learning cycles (e.g., Does the size of the planted clove influence the number of cloves in the eventual bulb?, How does the variety of garlic correlate with the size of a clove?). These questions provide opportunities for students to take more responsibility for their learning by challenging them to identify variables, form hypotheses, design experiments, and make predictions based upon their hypotheses. These opportunities give students a chance to do science.

Once the students understood the context, we handed each student a bulb of garlic, and gave them about $20$ minutes to study the bulb and work on these problems.

The specific garlic we used in class was a red softneck, of undetermined origin.  In general, the size of the final garlic bulb in July is proportional to clove size when planted the previous September, so not all of the cloves in a given bulb would be worth putting in a row.  Our cutoff was that only cloves with diameter similar to or larger than a dime would have an acceptable yield.  This means that by mass, only $91\%$ of the bulb was usable for seed. The bulbs averaged $15.5$ plantable cloves per bulb, with an average mass per plantable clove of about $3.5$ grams.

What follows are the specific questions we worked through with the students, common student misconceptions, the class data set, and solutions.

In order to answer these questions, students massed the bulbs, broke the bulbs into individual cloves, and decided what fraction of a given bulb would actually be useful for seed stock or eating. When these data were collected, we had each student write the measurements for his or her individual bulb on the class whiteboard. This data set is shown in Table \ref{table:student_data}. A typical bulb, and a bulb broken into cloves is shown in Figure \ref{fig:garlic_cloves}.

\begin{table}
%\begin{tabular}{cccccc}
\begin{tabular}{llllll}
\hline
Student 	&	\textrm{Bulb} 		&\multicolumn{2}{l}{Plantable Cloves}  & \multicolumn{2}{l}{Unplantable Cloves} \\
		&	\textrm{Mass} (g)	& Number	& \textrm{Mass} (g)		& Number & Mass (g)\\
\hline
1	& 57.152	&16	&54.832	&0	&2.32\\
2	& 70.757	&16	&62.632	&5	&8.125\\
3	& 51.712	&17	&49.059	&1	&2.653\\
4	& 55.136	&16	&50.573	&2	&4.563\\
5	& 57.214	&17	&46.220	&7	&10.994\\
6	& 63.071	&15	&54.421	&4	&8.65\\
7	& 60		&16	&58.18	&0	&1.82\\
8	& 53.326	&14	&49.086	&4	&4.24\\
9	& 59.2	&15	&56.97	&0	&2.23\\
10	& 66.89	&12	&61.956	&5	&4.934\\
11	& 61.29	&14	&53.33	&3	&7.96\\
12	& 59.568	&16	&57.986	&2	&1.582\\
13	& 71.088	&14	&59.967	&4	&11.121\\
14	& 62.299	&13	&49.193	&6	&13.106\\
15	& 60.438	& 21	&58.751	&7	&1.687\\
\hline
average	&60.609	&15.467	&54.877	&3.333	&5.732\\
median	&60.000	&16		&54.832	&4		&4.563\\
st. dev.	&5.674	&2.100	&5.154	&2.440	&3.935\\
\hline
\end{tabular}
\caption{Student data as it was written on the board in class.   The bottom rows of the table show the averages taken over the class.  The fluctuations in the reported precision of mass measurements led to a discussion of significant figures.}
\label{table:student_data}
\end{table}

\subsection {How much garlic should I purchase to have adequate seed?}

In an average week, I use up about 1 bulb of garlic. There are $52$ weeks/year, so I need to have about $52$ bulbs of garlic available to have supply through the year. Some garlic will undoubtedly be lost to spoilage. If we estimate the amount lost to be $20\%$ of the total (so $80\%$ of the harvest will be usable) then we can calculate the portion needed for harvest with the proportion,

\be
\frac{\textrm{Garlic~not~spoiled}}{\textrm{Garlic~harvested}} =\frac{80\%}{100\%}=0.8. \nonumber
\ee

This implies that we'll need $52 bulbs/0.8 = 65~bulbs$.  To have seed to plant in the following fall, we should also plan to have more than $65$ additional cloves to use for seed.  Spoilage between harvest in June and planting in September will be lower than that during the whole year. If we estimate $90\%$ viability after $3$ months, then, using similar logic as above, to plan next year's crop we'll need about $72$ additional cloves at harvest time.

Student data indicated that there were about $15.5$ usable cloves of garlic per bulb, so $5$ extra bulbs, for a total of $70$ bulbs, would probably provide sufficient garlic for a year's consumption, planting, and loss.

\subsection{How big an area do I need to cultivate?}

In addition to the planting description already provided, we told the students that the planned garlic patch was tentatively a $2$ foot by $15$ foot section of soil.  If we assume the rows run the ``long'' way down the patch, and that the bulbs are planted every $4$ inches along the row, a person can use dimensional analysis to calculate the number of cloves that can be planted in a single row:

\be
\frac{15~\textrm{feet}}{\textrm{row}}
	\frac{12~\textrm{inches}}{\textrm{foot}} 
	\frac{1~\textrm{clove}}{4~\textrm{inches}} 
	= 45 \frac{\textrm{cloves}}{\textrm{row}}\nonumber
\ee

At this point in the student discussion there is a significant opportunity to connect the proportional reasoning to the actual ``common-sense'' layout of the garden. The number of rows that can be planted in a $2$ foot width is determined by the way the rows are laid out.  If, for example, the outside rows run at the absolute edge of the bed, one could fit $5$ rows of garlic in the bed.  However, horticulturally speaking, the foot traffic and lack of cultivation at the edge of the bed would probably lead to a significantly lower yield in the two outside rows, and better yields might actually be seen if $4$ rows are planted, with a $3$ inch buffer on either side of the outside rows.

If we use $4$ rows, with $45$ cloves per row, the patch could hold about $180$ garlic plants, which is probably more than enough for a family's annual needs.

\subsection{How much money will I save by buying and growing plants, rather than buying bulbs of equivalent quality at the supermarket?}

Looking at the data, we see that the average bulb of garlic has a mass of $60.6$ grams.  If we want to splurge and plant $180$ cloves (the maximum that our garlic patch will allow), the mass of garlic I need to buy at the store can be calculated with dimensional analysis,

\bea
	\left(180~\textrm{cloves}\right)
	\left(\frac{1~\textrm{bulb}}{15.5~\textrm{plantable~cloves}}\right)\times \nonumber\\
	\times \left(60.6\frac{\textrm{grams}}{\textrm{bulb}}\right)
	\left( \frac	{1~\textrm{lb}}{454~\textrm{grams}}\right) 
	= 1.5 \textrm{lbs~of~garlic} \nonumber
\eea

Locally, garlic sells for $\$2.69$ per pound, so our seed would cost about $\$4.15$.  If $90\%$ of the plants survived and $20\%$ of the garlic was lost in storage, the $180$ plants would become $180\times0.9\times0.8=129$ bulbs of garlic.  If we assume that this garlic is similar to the seed stock, the mass of $129$ bulbs in the predicted harvest would be,

\be
129~\textrm{bulbs}\left(60.6\frac{~\textrm{grams}}{~\textrm{bulb}}\right)\left(\frac{1\textrm{lb}}{454~\textrm{grams}}\right) = 17.2~\textrm{lbs}
\ee

A similar calculation for the (grocery store) garlic a family might buy and use in a year is,

\be
\left(52\frac{\textrm{bulbs}}{\textrm{year}}\right)\left(60.6\frac{\textrm{grams}}{\textrm{bulb}}\right)\left(\frac{1~\textrm{lb}}{454~\textrm{grams}}\right) = 6.9~\textrm{lbs}
\ee

There are several things to note about this calculation. First, at grocery store prices, $\$2.69/lb$, $1$ bulb a week from the grocery store is about $\$18.70$ worth of garlic over a year, so it is slightly more profitable to grow your own garlic.  Second, the organic ``heirloom'' garlic at my local food co-op sells for upwards of $\$4/lb$, and the stuff a person produces in their backyard can certainly measure up to this quality standard.  If you were to buy the postulated $17.2$ lbs that the garlic patch produces, you'd be looking at a nearly $\$70$ cost, so if food quality is your concern, it is absurd not to grow your own garlic. There are a number of similar comparisons, like cost per seed, average yield per seed, etc, along these lines that can be performed.

\section{Assessment}

When most of the students were nearly finished with these problems, we presented them with the following simple question. ``I have a few bulbs of garlic left over that haven't been opened.  How many cloves would you guess are in one of those bulbs, and about how much do you think that clove would weigh?''

As mentioned earlier, the goals of this class include developing students' scientific thinking skills as well as their content knowledge. Therefore, the question ``How many cloves are in the next bulb?'' is critical because it challenges students to apply probabilistic reasoning, a scientific thinking skill that is often neglected in physical science classrooms. Many students used the class average of cloves/bulb to predict the number of cloves in the next bulb. However, we asked the students to describe a reasonable number or range of cloves to expect, which forced the class to be more specific in their rationale for a given prediction.

To come to collective agreement, we summarized the students' discussion by creating the histogram shown in Figure 3.  To make this figure, we computed how many times an exact number of plantable cloves was found in a bulb of garlic.  As the data shows, the most common (and we argued, most probable) outcome was 16 plantable cloves. In talking about a range of possible outcomes, one can then say that 5 times out of 15, or 1/3 of the time, students in the class saw this outcome.  Further, if the range is 15-17 plantable cloves, this accounts for 9 of the 15 outcomes, which is almost 2/3 of the total probability.  Finally, the range of 14-17 plantable cloves accounts for 12 of the 15 outcomes, which is $80\%$ of the total probability.  With this sort of building logical argument, it isn't hard to transition into the language of being ``$80\%$ confident'' that the next clove would contain 14-17 cloves of garlic.  Fortunately for us, the two remaining bulbs contained respectively 16 and 15 plantable cloves, which seemed to further strengthen the power of the statistical argument in the our students' minds.

\begin{center}
\begin{figure}
\caption{To summarize the students' observations, the class constructed a histogram of plantable cloves per bulb of garlic.  This figure was used to explain the meaning of average number of plantable cloves seen per bulb ($15.5$ cloves), and the standard deviation about that average ($2.1$).  Further discussion of the data led to the introduction of confidence intervals, ie, the probability that a new bulb of garlic has a certain number of cloves.  For example, since $12$ of the $15$ total outcomes correspond to $14$ to $17$ plantable cloves per bulb of garlic, we can say with probability of $12/15 = 0.75$, or with $75\%$ confidence, that another bulb of garlic from the same produce bin at the grocery store would contain $14$ to $17$ plantable cloves.
}
\includegraphics[width=\columnwidth]{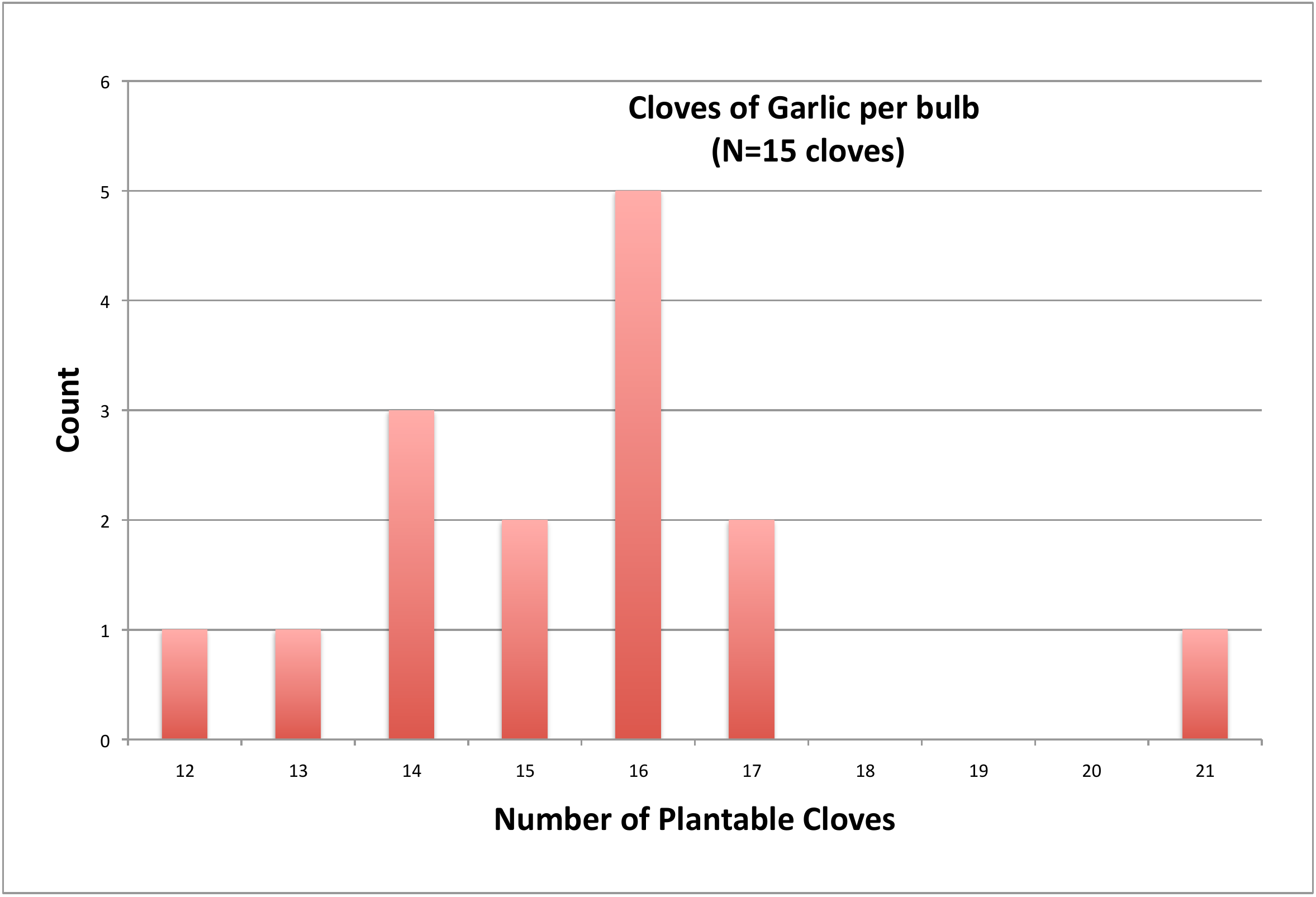}
\label{fig:garlic_histogram}
\end{figure}
\end{center}
\section{Conclusions}

Challenging students to apply course concepts and thinking skills in new contexts is fundamentally important.  Speaking plainly, if our students can't apply what theyÕve learned to a new situation, it becomes difficult to say that they learned anything.  Our rationale for taking one class period of chemistry and devoting it to a ``non-chemistry'' topic is that our goal (described in Deming and Cracolice, 2004) is to teach students how to think, rather than teach them what to think. Therefore, providing students with a less-familiar or far-transfer type problem to solve, challenges them to take data and make sense of those data in order to arrive at a plausible answer. These are the skills necessary to become competitive in college science classrooms, which warrant a certain amount of educational time, even at the expense of a few class periods of traditional content.

\section{Acknowledgements}
This work was supported in part by the Minnesota State Colleges and University SystemÕs Center for Teaching and Learning Instructional Development Grant Program.

\end{document}